\documentclass[runningheads]{llncs}
\usepackage[T1]{fontenc}
\usepackage{graphicx}
\usepackage{url}
\usepackage{listings}
\usepackage{xcolor}
\usepackage{cleveref}

\lstdefinelanguage{rdfmessages}{
  morekeywords={@prefix,@base,@version,PREFIX,BASE,VERSION,MESSAGE,a},
  sensitive=true,
  morecomment=[l]{\#},
  morestring=[b]"
}
\lstdefinelanguage{turtle}{
  morekeywords={@prefix,@base,a},
  sensitive=true,
  morecomment=[l]{\#},
  morestring=[b]"
}
\begin{document}
\title{It’s Time to Standardize RDF Messages}
%
%
\author{Pieter Colpaert\inst{1}\orcidID{0000-0001-6917-2167} \and
Piotr Sowiński\inst{2,3}\orcidID{0000-0002-2543-9461}}
\authorrunning{Colpaert and Sowiński}
%
\institute{IDLab -- UGent -- imec\\
\email{pieter.colpaert@ugent.be}\and
NeverBlink, ul. Wspólna 56, 00-684 Warsaw, Poland\and
Warsaw University of Technology, pl. Politechniki 1, 00-661 Warsaw, Poland\\
\email{piotr@neverblink.eu}}
\maketitle              
\begin{abstract}
RDF-based systems increasingly operate in event-driven and streaming settings, where producers and consumers exchange data as discrete units of communication rather than as freely mergeable RDF statements.
As existing RDF semantics and tooling do not provide an interoperable notion of what statements belong together as one message, developers often rely on out-of-standard techniques, transport-level assumptions, or heuristics, leading to interoperability problems and inefficiencies.
We propose the concept of an RDF Message as an RDF Dataset intended to be interpreted atomically as a single communicative act, laying the foundation for defining RDF Message Streams and RDF Message Logs. 
The proposal makes message boundaries explicit across serializations, transport, and storage systems, which in turn enables incremental consumption and reproducible replay in use cases such as IoT observations, archived RDF streams, nanopublications, or processing SPARQL CONSTRUCT results.
Building on this, RDF Message Profiles, such as Linked Data Event Streams or ActivityStreams, then provide the terms for describing pagination, message structure, ordering, or retention policies.
As part of the W3C Community Group on RDF Stream Processing, we are now seeking broader support and comments on the proposal from the Semantic Web community.

\end{abstract}

\section{Introduction}

The deployment context of RDF systems has shifted beyond static publication and batch-oriented querying.
Contemporary systems operate in event-driven and streaming settings, where information is produced, transmitted, and consumed continuously.
Examples include IoT sensor networks, activity feeds (cfr.\ ActivityPub), knowledge graph updates, benchmark stream distribution~\cite{riverbench}, Trustflows~\cite{trustflows-1}, Nanopublications~\cite{nanopubs}, or returning multiple results when evaluating SPARQL CONSTRUCT queries.
In these environments, we need a standard understanding of how RDF statements are grouped as units of communication.

A persistent practical issue is that RDF’s standard semantics and its mainstream processing ecosystem are largely agnostic to message boundaries.
RDF graphs and datasets are typically treated as collections of statements that can be merged, reordered, and partitioned without semantic effect.
This is appropriate for many knowledge representation tasks, but it becomes problematic when the producer intends a specific grouping of triples or quads to be interpreted atomically as a single communicative act, for instance: ``this set of statements constitutes one observation'', ``this is one update in a stream'', or ``this is one query result''.
When explicit boundaries are absent, consumers often rely on transport-level assumptions, such as treating newlines as delimiters or using a streaming protocol such as WebSockets where messages are already present at the transport layer.
Alternatively, consumers may apply heuristics to reconstruct the intended message, such as Concise Bounded Description\footnote{\url{https://www.w3.org/submissions/CBD/}} or ad hoc uses of vocabulary constructs and storage features.
Such approaches are non-standard patchwork that will break once messages need to travel across multiple systems and different sets of tooling.

A commonly suggested workaround is to use named graphs as the grouping mechanism.
However, this conflates two separate concerns: scoping statements for dataset semantics versus grouping statements for communication.
Named graphs provide a way to partition statements \emph{within} an RDF dataset, whereas messages capture what a producer intends to communicate \emph{as one exchange unit}.
For example, a single message may contain statements in multiple named graphs and in the default graph; conversely, a message may even be empty, for example to represent a heartbeat or keep-alive event.
We illustrate this in \cref{message}, which replicates how named graphs are used in event-oriented streaming Linked Data systems~\cite{nanopubs,ldes,bonte2025languages,dedecker2025demonstrating,mauri2016triplewave}.

\begin{lstlisting}[language=turtle,label=message,caption={One RDF Message may contain statements in the default graph and in named graphs. The message boundary should not be conflated with the boundary of a named graph.}]
_:b0 prov:generatedAtTime "2026-01-30T10:05:34Z"^^xsd:dateTime .
_:b0 {  ex:Observation1 a sosa:Observation ;
          sosa:resultTime "2026-01-30T09:52:30Z"^^xsd:dateTime ;
          sosa:hasSimpleResult 21.4 . }
\end{lstlisting}

Either way, named graphs also do not solve a performance problem.
In RDF, the order of RDF statements does not carry semantic meaning.
As a consequence, statements belonging to the same named graph or message may appear far apart in a serialization, and extracting one complete graph or message may require parsing the full input first. 

In this paper, we argue that message grouping must be made explicit as a first-class interoperability concern for RDF-based streaming and event-driven systems.
The next section introduces our proposal of RDF Messages\footnote{The RDF Messages specification is available online at \url{https://w3c-cg.github.io/rsp/spec/messages} and links an issue board where discussions take place.}, a draft report in the RDF Stream Processing Community Group.

\section{RDF Messages}

An \emph{RDF Message} is an RDF Dataset that is intended to be interpreted atomically as \textbf{a single communicative act}.
The concept does not change the RDF semantics of the statements inside the dataset.
Rather, it makes explicit that a particular grouping of statements belongs together as one message for communication, storage, and processing purposes.
An RDF Message may also be empty, for example to represent a keep-alive event.

Building on this concept, we define an \textbf{RDF Message Stream} as an ordered sequence of RDF Messages, potentially unbounded, transmitted from one system to another.
Furthermore, we also define an \textbf{RDF Message Log} as a replayable representation of such a stream that preserves message order and boundaries.
A stream is consumed as it is produced, whereas a log allows the same sequence of messages to be archived, exchanged, replayed, or processed again later.

By default, each RDF Message is only asserted in \textbf{its own context}.
Therefore, RDF Messages should not automatically be interpreted as if all statements from all messages were asserted in one global dataset.
Instead, each message forms its own unit of communication and interpretation.
A consumer or stream processor may derive downstream assertions from a message, but that step depends on the application and its operational setting.
This keeps the core concept of RDF Messages compatible with existing RDF semantics while avoiding the ambiguity that arises when messages are implicitly merged.
For example, one message may indicate that the radio is playing song A, while another message indicates that the same radio is playing song B. 
This is not a contradiction, as these messages refer to different contexts, i.e., different moments in time.

\textbf{Blank node identifiers} in RDF Message Streams and RDF Message Logs are scoped to the message in which they occur.
This allows processors to handle long-running streams incrementally, without having to preserve blank node identifiers across message boundaries or risk identifier collisions.
When applications need to refer to the same resource across multiple messages, they can still use skolemization.

The core notion of RDF Messages is intentionally minimal: it does not prescribe one fixed way to identify messages, attach timestamps, express delivery semantics, or define how message contents should affect downstream state.
These aspects depend on the application domain and are therefore better specified separately.
We foresee \textbf{RDF Message Profiles} to define how messages can be structured and interpreted in a given setting.
Such profiles may constrain the shape and ordering of messages, define how the chronology of a stream is determined, specify versioning rules, define hypermedia controls for paginating a log, describe transaction boundaries, define how changes are to be applied, or indicate retention policies.
These are features that can already be expressed using Linked Data Event Streams~\cite{ldes}, and can be partially supported by specifications like PROV-O, ActivityStreams, SSN/SOSA, or SAREF.
For the example in \cref{message}, we could define a PROV-O RDF Message profile specifying that the \texttt{prov:generatedAtTime} defines the chronological order, and maybe even configure that \texttt{sosa:resultTime} defines the version order.

RDF Messages do not by themselves prescribe a concrete syntax.
To exchange, archive, or replay RDF Messages, \textbf{serializations} need a way to preserve message boundaries explicitly.
RDF 1.2 introduces version announcement mechanisms, which we use as a hook to indicate that a document uses message-aware syntax.
This allows a parser to know from the start that it should preserve message boundaries while reading the input.
\Cref{messagelog} illustrates the idea for a Turtle-like RDF Message Log.
The document starts by announcing a message-aware version.
The \texttt{MESSAGE} delimiter then marks the boundary between successive RDF Messages.
This way, a parser can emit complete messages incrementally as soon as a delimiter is encountered, instead of buffering an entire file and guessing where one intended unit ends and the next begins.
This also allows an empty message to be represented explicitly.

\begin{lstlisting}[language=rdfmessages,label=messagelog,caption={Example of a Message Log in Turtle with explicit message boundaries. The second message is empty, and the blank node label \texttt{\_:b0} is reused in the first and third messages because blank node identifiers are scoped per message.}]
VERSION "1.2-messages" # Prefixes omitted
_:b0 sosa:resultTime "2026-05-12T18:20:00Z"^^xsd:dateTime ;
    sosa:hasSimpleResult 22 .
MESSAGE # an empty message, e.g., a heartbeat
MESSAGE # the third message is another observation
_:b0 sosa:resultTime "2026-05-12T18:25:00Z"^^xsd:dateTime ;
    sosa:hasSimpleResult 23 .
\end{lstlisting}

This is only one way to serialize RDF Message Logs.
We also support RDF Message adoption proposal in other serializations, such as YAML-LD, JSON-LD, or Jelly-RDF~\cite{sowinski2025jelly} that already has the notion of frames; and ecosystems, such as types in the RDF-JS TypeScript ecosystem. 

\section{Use Cases}

\emph{Internet of Things.}
IoT devices emit streams of discrete messages, e.g., temperature observations from a thermometer. While ontologies like SOSA/SSN allow one to describe such observations, the messaging mechanism remains undefined. RDF Message Streams provide the interoperability basis for such streams, needed for unambiguous interpretation of IoT data.

\emph{Event-driven knowledge graph updates.}
Microservices frequently emit ``patch-like'' RDF payloads representing updates. RDF Messages can represent each update dataset as an atomic message, while allowing deployments to choose whether the update semantics are interpreted as INSERT/DELETE operations (out-of-band) or as domain-level events (in-band). A real-life example of this is the Nanopublication Network~\cite{nanopubs}, which continuously replicates streams of new Nanopublications (each being an RDF Dataset) across services.

\emph{SPARQL CONSTRUCT results as units.}
When answering a SPARQL CONSTRUCT query, SPARQL endpoints return the triples that can be used by a consumer to reconstruct the answers.
We propose to use RDF Messages, where the consumer can simply understand from the serialization which triples belong to one specific result. This enables more granular reuse of SPARQL CONSTRUCT results in query clients.

\emph{Archiving and replaying of RDF streams.}
Archiving continuous RDF output as a log is common for analytics, debugging, and reproducibility. Without message boundaries, archives become ambiguous: replay may emit different segmentations than original, affecting downstream consumers that rely on atomic events. RDF Message Logs address this by storing each dataset as a record with an explicit boundary and optional offset/time metadata.

\section{Conclusion}

RDF Messages offer a simple and standardizable abstraction, while leaving use case-specific interpretation to profiles built on top.
As this proposal is being developed within the W3C Community Group on RDF Stream Processing, we now invite comments, implementations, participation, and broader community support to help refine the specification and move toward convergence on a shared approach for RDF Messages. The current specification is available at \url{https://w3c-cg.github.io/rsp/spec/messages}.

\begin{credits}
\subsubsection{\ackname} Pieter Colpaert’s work was funded by Serendipity Engine (Research Foundation -- Flanders (FWO) grant number S006323N).
\end{credits}
\bibliographystyle{splncs04}
\bibliography{references}
\end{document}